\documentclass[aps,twocolumn,superscriptaddress,prd,nofootinbib]{revtex4}

\usepackage{graphicx}
\usepackage{url}
\usepackage{multirow}
\usepackage{amsmath}
\usepackage{amssymb}
\usepackage{amsfonts}
\usepackage[colorlinks,linkcolor=blue,anchorcolor=blue,citecolor=red]{hyperref}
\usepackage{xcolor}
\usepackage{acronym}
\usepackage{CJK}
\usepackage{orcidlink}

\newcommand{\TRC}{MOE Key Laboratory of TianQin Mission, TianQin Research Center for Gravitational Physics \& School of Physics and Astronomy, Frontiers Science Center for TianQin, Gravitational Wave Research Center of CNSA, Sun Yat-sen University (Zhuhai Campus), Zhuhai 519082, China}

\begin{document}
\begin{CJK*}{UTF8}{gbsn}

\title{Space-based Gravitational Wave Observatories Will Be Able to Use Eccentricity to Unveil Stellar-mass Binary Black Hole Formation}

\author{Han Wang(王晗)\orcidlink{0009-0007-5095-9227}}
\affiliation{\TRC}

\author{Ian Harry\orcidlink{0000-0002-5304-9372}}
\affiliation{University of Portsmouth, Portsmouth, PO1 3FX, United Kingdom}

\author{Alexander Nitz\orcidlink{0000-0002-1850-4587}}
\affiliation{Department of Physics, Syracuse University, Syracuse NY 13244, USA}
\affiliation{Max-Planck-Institut f\"ur Gravitationsphysik (Albert-Einstein-Institut), D-30167 Hannover, Germany}

\author{Yi-Ming Hu(胡一鸣)\orcidlink{0000-0002-7869-0174}}
\email{huyiming@mail.sysu.edu.cn}
\affiliation{\TRC}

\date{\today}

\begin{abstract}
The measurement of eccentricity would provide strong constraints on the formation channel of stellar-mass binary black holes.
However, current ground-based gravitational wave detectors will, in most cases, not be able to measure eccentricity due to orbital circularization.
Space-based observatories, in contrast, can determine binary eccentricity at 0.01Hz to $e_{0.01}\gtrsim\mathcal{O}(10^{-4}) $.
Directly observing stellar-mass binary black holes with space-based observatories remains a challenging problem. However, observing such systems
with ground-based detectors allows the possibility to identify the same signal in archival data from space-based observatories in the years previous.
Since ground-based detectors provide little constraints on eccentricity, including eccentricity in the archival search will increase the required number of filter
waveforms for the archival search by 5 orders of magnitudes [from $\sim \mathcal{O}(10^3)$ to $\sim \mathcal{O}(10^8)$], and will correspondingly need $ \sim8\times10^5 $ core hours (and $ \sim 10^5$ GB of memory), even for a mild upper limit on eccentricity of $0.1$.
In this work, we have constructed the first template bank for an archival search of space-based gravitational wave detectors, including eccentricity. 
We have demonstrated that even though the inclusion of eccentricity brings extra computational burden, an archival search including eccentricity will be feasible in the time frame of planned space-based observatories, and will provide strong constraints on the eccentricities of stellar-mass binary black holes.
\end{abstract}

\maketitle
\end{CJK*}

\acrodef{GW}{gravitational wave}
\acrodef{sBBH}{stellar-mass binary black hole}
\acrodef{EM}{electromagnetic}
\acrodef{SNR}{signal-to-noise ratio}
\acrodef{FIM}{Fisher information matrix}
\acrodef{MCMC}{Markov Chain Monte Carlo}
\acrodef{PSD}{power spectral density}
\acrodef{LISA}{Laser Interferometer Space Antenna}
\acrodef{PN}{post-Newtonian}
\acrodef{TDI}{time delay interferometry}
\acrodef{SPA}{stationary phase approximation}
\acrodef{AGN}{active galactic nuclei}

\section{Introduction}
Stellar-mass black holes (sBBHs) detected before 2015 were mainly observed through X-ray binaries ~\cite{Orosz2011,CorralSantana2016}, with measured masses  $\lesssim 20 M_\odot$ ~\cite{Oezel2010}. 
The first \ac{GW} signal GW150914 observed by LIGO and Virgo has been identified as the coalescence of \ac{sBBH} with component masses $36_{-4}^{+5}M_\odot$ and $29_{-4}^{+4}M_\odot$~\cite{Abbott2016}.
The observed masses posed a significant challenge to our understanding of the formation mechanism of \acp{sBBH}~\cite{Abbott2016b}.
To date, nearly 100 \ac{sBBH} mergers have been reported, many of them as heavy as GW150914~\cite{Abbott2021,Nitz2021}.
With the accumulation of \ac{GW} observations, numerous models have been proposed to explain the formation of these \acp{sBBH}~\cite{Barack2019}.
The eccentricity of a \ac{sBBH} system is a key probe in unveiling the system's formation mechanism. 
However, among all \ac{GW} detections, none was claimed to have measurable eccentricity (eccentricity at 10 Hz $ e_{10} \gtrsim 0.1 $)~\cite{Abbott2019,RomeroShaw2019,Wu2020} until GW190521, which some argue could be eccentric~\cite{Abbott2020c,RomeroShaw2020,Gayathri2022}. The sensitive frequency band of current ground-based detectors makes them only capable of observing \acp{sBBH} seconds before coalescence. 
Advanced LIGO/Virgo can measure the eccentricity for binaries with $ e_{10} \gtrsim 0.05 $~\cite{Lower2018}, but most \acp{sBBH} cannot retain eccentricity that high
because of orbital circularization due to gravitational wave emission before entering the ground-based frequency band~\cite{Peters1964}. 
Therefore, it is challenging for ground-based detectors to distinguish and identify the formation channels of \acp{sBBH}~\cite{Rodriguez2018}.

Space-based \ac{GW} observatories, like TianQin~\cite{Luo2016} and \ac{LISA}~\cite{AmaroSeoane2017}, offer a promising solution to this question. 
They have longer baselines than their ground-based counterparts and therefore are sensitive in a lower frequency band and could observe \acp{sBBH} for years. 
This makes space-based observatories capable of precise mass measurements and unveiling the evolution of eccentricity and spin of \ac{sBBH} sources~\cite{Nishizawa2016,Nishizawa2017,Chen2017,Liu2020,Buscicchio2021,Klein2022}. For example, eccentricity evolves as $ e \sim e_\mathrm{i} (f/f_\mathrm{i})^{-19/18} $ at leading order~\cite{Yunes2009}. If the \ac{GW} of a binary system evolves to the ground-based detector frequency band at $ f \gtrsim 1 \mathrm{Hz} $ with eccentricity equal to $ 10^{-3} $, the system has a significantly larger eccentricity, $ e_\mathrm{i}\sim0.1 $, at a frequency $ f_\mathrm{i}\sim0.01 \mathrm{Hz} $, which is a typical sensitive frequency for space-based observatories.

\begin{figure}[ht]
	\centering
	\includegraphics[width=\linewidth]{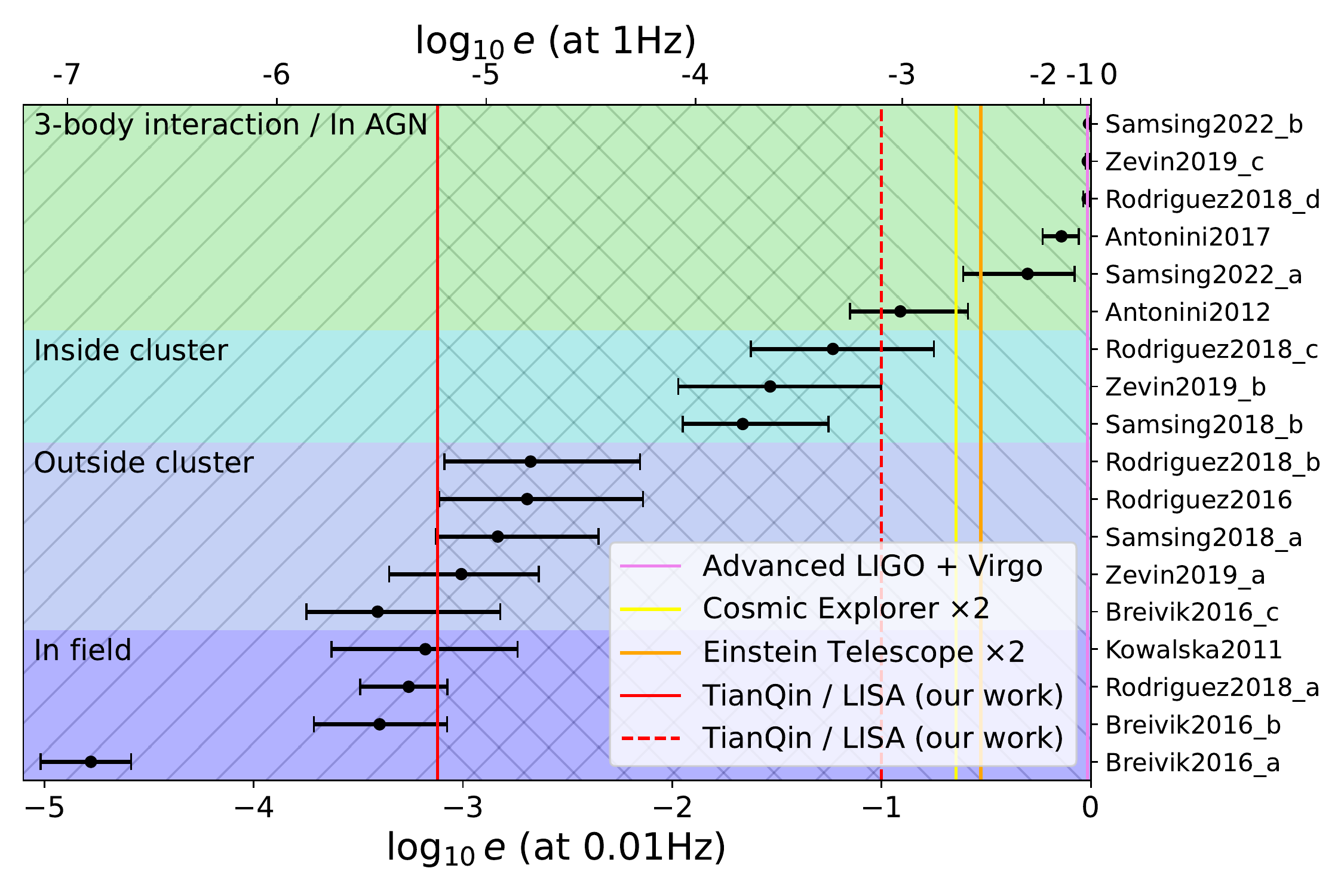}
	\caption{Predicted eccentricity distributions from different evolution models.
        The black dots and error bars represent the median values and 50\% credible intervals, respectively. 
    The vertical solid (dashed) lines indicate the minimum (maximum) detectable  eccentricities of different \ac{GW} observatories.
    }
	\label{citation_ecc}
\end{figure}

Figure \ref{citation_ecc} shows eccentricity distributions predicted by different evolution models. \acp{sBBH} formed in isolation are likely to have $e_{0.01}\lesssim10^{-3}$~\cite{Kowalska2011,Breivik2016}. 
\acp{sBBH} dynamically formed in globular clusters and subsequently ejected into the field have similar distributions, with $e_{0.01}\lesssim10^{-2}$~\cite{Breivik2016,Rodriguez2016}. 
However, \acp{sBBH} that evolve inside clusters can retain a high eccentricity with $e_{0.01}\gtrsim10^{-2}$~\cite{Samsing2018}, and eccentricities can reach extreme values ($e_{0.01}\sim1$) for systems involved in various triplets~\cite{Antonini2012,Antonini2017,Zevin2019} or in \ac{AGN} disks~\cite{Samsing2022}. 
Space-based observatories have the capability to detect eccentricities $e_{0.01}\gtrsim10^{-3}$~\cite{Nishizawa2016,Liu2020}.
Therefore, \ac{sBBH} detections with space-based observatories, alongside observations with ground-based facilities, offer a unique opportunity to identify the formation channel of \acp{sBBH}.

Considering eccentricity for the \acp{sBBH} can bring additional benefits.
The inclusion of eccentricity can break parameter degeneracy~\cite{Xuan2023},
improve the precision of measuring source distance and sky localization~\cite{Sun2015,Yang2022},
and make future tests of general relativity more reliable~\cite{Saini2022,Bhat2023}.

Matched-filtering methods have been widely used in ground-based \ac{GW} detection~\cite{Babak2013}. These searches require a suitable set of waveform filters, or ``template bank''. Applying this method to TianQin or LISA will be challenging because of the number of waveform templates required. An example search for compact binary mergers in LIGO/Virgo data requires $ \lesssim 4\times10^{5} $ templates~\cite{DalCanton2017}. In contrast~\citet{Moore2019a} predicts that a bank of order $ 10^{30} $ templates would be needed to cover the whole \ac{sBBH} parameter space for LISA, far exceeding a reasonable computational cost. 

It has been proposed that a search of archival data from space-based observatories, triggered by detection with ground-based facilities, can achieve the multiband detection of \acp{sBBH}~\cite{Sesana2016,Wong2018,Ewing2021,Klein2022}.
Next-generation ground-based detectors, like Einstein Telescope (ET)~\cite{Punturo2010} and Cosmic Explorer (CE)~\cite{Reitze2019} will be able to detect \ac{GW} events with \acp{SNR} $\mathcal{O}(10^{2-3})$ and will therefore place tight constraints on the source parameters, for example measuring the chirp mass to one part in $10^{6}$~\cite{Cho2022}. 
Therefore, the parameter space of an archival search of TianQin/LISA data can be greatly reduced and the required template bank size reduced to the level of $10^4$ templates~\cite{Ewing2021}.

However, the impact of the eccentricity on archival searches has not been explored. In this paper, for the first time, we implement a matched-filtering bank generation process for an archival search in space-based observatories incorporating eccentricity, triggered by an observation using next-generation ground-based detectors. 
Using GW150914 and GW190521 as examples, we find that even though the inclusion of eccentricity would enlarge the template bank by a factor of $ \sim\mathcal{O}(10^5) $, the task is still tangible. 
This work provides a practical solution to the realistic multiband \ac{GW} observation scenario.

\section{Methodology}
To detect \acp{GW} by matched filtering, we use \texttt{EccentricFD}~\cite{Yunes2009,Huerta2014}, a nonspinning inspiral-only frequency-domain waveform approximant with eccentricity at the initial frequency $e_\mathrm{i}$ valid up to 0.4, 
for constructing the template bank. \texttt{EccentricFD} includes \ac{PN} corrections up to 3.5PN order and has been included into \texttt{LALSuite}~\cite{LSC2018}. The eccentricity in \texttt{EccentricFD} is expanded to $ \mathcal{O}(e^8) $ and then further expanded in $ e_\mathrm{i} $ up to $ \mathcal{O}(e_\mathrm{i}^8) $.
The parameter set follows $ \lambda^\mu = \left(\mathcal{M},\eta,D_L,t_c,\phi_c,\iota,\lambda,\beta,\psi,e_\mathrm{i} \right) $, where $ \mathcal{M} \equiv {\left( {{m_1}{m_2}} \right)^{3/5}}{\left( {{m_1} + {m_2}} \right)^{ - 1/5}} $ and $ \eta  \equiv \left( {{m_1}{m_2}} \right){\left( {{m_1} + {m_2}} \right)^{ - 2}} $ given by the component masses $ m_1 $ and $ m_2(m_1>m_2) $ are the chirp mass and symmetric mass ratio, $ D_L $ is the luminosity distance, $ t_c $ and $ \phi_c $ are the coalescence time and phase, $ \iota $ is the inclination angle, $ (\lambda,\beta) $ are ecliptic longitude and ecliptic latitude, $ \psi $ is the polarization angle and $ e_\mathrm{i} $ is the eccentricity at the initial frequency $ f_\mathrm{i} $ in the quadrupolar \ac{GW} mode. For space-based observatories, $ f_\mathrm{i} $ is determined by the evolution time $T$ from the beginning of observation to the merger. In this work, we assume a fully continuous five-year observation for both TianQin and LISA, and the merger happens at the end of the five-year period. 
For $ M_\mathrm{tot} \lesssim 10^5 M_\odot $ and $ T \gtrsim 1 \mathrm{yr} $, the correction for $ f_\mathrm{i} $ from the eccentricity can be neglected(see Ref.~\cite{Yunes2009}, Appendix E), so we will use the noneccentric frequency-time relation at leading PN order in the following calculation: $ f_\mathrm{i}=(5 / 256)^{3 / 8} {\pi}^{-1} \mathcal{M}^{-5 / 8}T^{-3 / 8}$.

The size of the parameter space that would need to be searched in an archival search depends on the parameter estimation precision of the next-generation ground-based detectors. One can use the \ac{FIM} $ \Gamma_{i j}$ to estimate the statistical uncertainties in measuring parameters. $ \Gamma_{i j}=\left(\frac{\partial h}{\partial \lambda^{i}} \middle| \frac{\partial h}{\partial \lambda^{j}}\right) $, where $ (h | g)\equiv 4 \Re \int_0^{+\infty} \frac{\tilde{g}^*(f) \tilde{h}(f)}{S_n(f)} \mathrm{d} f $, $ S_n(f) $ is the one-sided detector noise power spectral density, $ \tilde{h}(f) = \tilde{h}(f, \lambda^\mu) $ is the Fourier transform of the waveform $ h(t) $, and $ \lambda^\mu $ is the parameter set. The overall \ac{FIM} of a detector network is the summation of the \ac{FIM} of each detector. Under the Gaussian stationary assumption, the covariance matrix can be approximated by $ \Sigma=\Gamma^{-1} $, and the marginalized parameter uncertainties can be estimated as $ {\sigma _{{\lambda ^i}}} = \sqrt {{\Sigma _{ii}}} $.

Here we consider a ground-based detector network including ET and two CEs, with  their sites randomly chosen.
Since \ac{GW} emission will cause a binary orbit to circularize over time~\cite{Peters1964}, we assume that events are noneccentric in the ground-based observation window. Higher-order modes, however, will be important for ET or CE, especially given the large \ac{SNR} that events visible to LISA and TianQin will have. We therefore use the noneccentric \texttt{IMRPhenomHM}~\cite{London2018} waveform to estimate the precision with which next-generation ground detectors can measure source parameters. 
We choose a low-frequency cutoff of $ f_\mathrm{low}=1 \mathrm{Hz} $ for both CE and ET during the calculation. This is motivated by the result that one would acquire ~20\% of the whole \ac{SNR} between 1 and 10Hz with ET~\cite{Lenon2021}.
Our estimation is consistent with previous studies~\cite{Ewing2021,Liu2020,Sesana2016}, which show that for a \ac{GW} event that retains no eccentricity when entering the ground-based observation window, the only two parameters that space-based observatories can measure more precisely are the chirp mass $\mathcal{M}$ and initial eccentricity $e_\mathrm{i}$. Therefore we assume that all the parameters except for chirp mass and eccentricity are known exactly when performing an archival search, and the chirp mass range is determined by the uncertainty from the network of the ET and two CEs, i.e.,\ $\mathcal{M} \in [\mathcal{M}_0-10\sigma_\mathcal{M},\mathcal{M}_0+10\sigma_\mathcal{M}]$. 
In the future, we should directly use the posterior from Bayesian inference in ground-based detectors, but for this study, the uncertainty range generated by the \ac{FIM} is a reasonable and conservative estimate.

We construct a template bank using \texttt{sbank}~\cite{sbank,Babak2008,Harry2009}, a \texttt{PYTHON} package for generating stochastic template banks for compact binaries. When generating template banks stochastically we need to determine how much those two waveforms overlap with each other. The fitting factor (FF) is used to define the maximum ``similarity'' between a given waveform and the best matching template in a bank:~\cite{Creighton2011}
\begin{equation}
  {\rm FF}\left( {{\lambda ^\mu }} \right) \equiv \mathop {\max }\limits_{\lambda ^{\mu '}} \frac{{\left( h({\lambda ^\mu }) \middle| h({\lambda ^{\mu '}}) \right)}}{{\sqrt {\left( h({\lambda ^\mu }) \middle| h({\lambda ^\mu }) \right)\left( h({\lambda ^{\mu '}}) \middle| h({\lambda ^{\mu '}}) \right)} }}.
\end{equation}
Here $ \lambda ^{\mu '} $ denotes the parameter set for a template in the bank, and $ \lambda ^\mu $ is the parameter set for the test waveform. For a template bank to be complete (or ``valid''), any \ac{GW} signal in its parameter range should have $ {\rm FF} \geq M $, where $ M $ is the minimal match. Here we set $ M=0.97 $, which is a commonly used value~\cite{DalCanton2017,Abbott2021,Nitz2021}.

Ground-based detectors observe the \ac{GW} signal over a period of only seconds before coalescence, so that the Doppler frequency modulation from the movement at Earth's orbit can be ignored. However, the long observation time and the orbital motion of space-based observatories make the response time dependent, and one must consider these time-dependent response terms during bank generation. Additionally, unlike ground-based detectors that have fixed arm lengths during operation, the relative spacecraft motion results in unequal arm lengths. The method of \ac{TDI} has been proposed for canceling out the laser phase noise from different arms. It constructs particular combinations to make virtual equal arm interferometers. This is further complicated when considering eccentric waveforms. Here we use the frequency-domain \ac{TDI} response \cite{Marsat2021,Lyu2023} and combine it with \texttt{EccentricFD} which contains a set of eccentric harmonics. We follow the arm length and noise budget in~\citet{Luo2016} for TianQin, and $ L=2.5\times10^9\mathrm{m} $ with noise budget from~\citet{Babak2021} for LISA. We consider the response in the A channel as an example during all the calculations in this work.

Since different eccentric harmonics have different correspondences with the Fourier frequency, we should provide a frequency cutoff during the calculation to avoid the waveform generation exceeding the valid range for a specific \ac{GW} detector: $ \tilde h_\mathrm{det} = \sum_{j}\tilde h_j \times \Theta \left( j \cdot f_\text{high} - 2f \right)\Theta \left( 2f - j \cdot f_\text{low} \right)$, where $ \Theta(x) $ is the Heaviside step function and $ j $ denotes the $j$th eccentric harmonic \cite{Yunes2009}. For TianQin or LISA, we have
$ f_\text{low} = \max \left[ 10^{-4}\mathrm{Hz},f_0 \right], f_\text{high} = \min \left[ f_\text{ISCO},1\mathrm{Hz} \right] $, where $ f_\text{ISCO} = (6^{3/2} \pi (m_1+m_2))^{-1} $ is the quadrupolar frequency at innermost-stable circular orbit (ISCO).

\begin{figure}[ht]
	\centering
	\includegraphics[width=\linewidth]{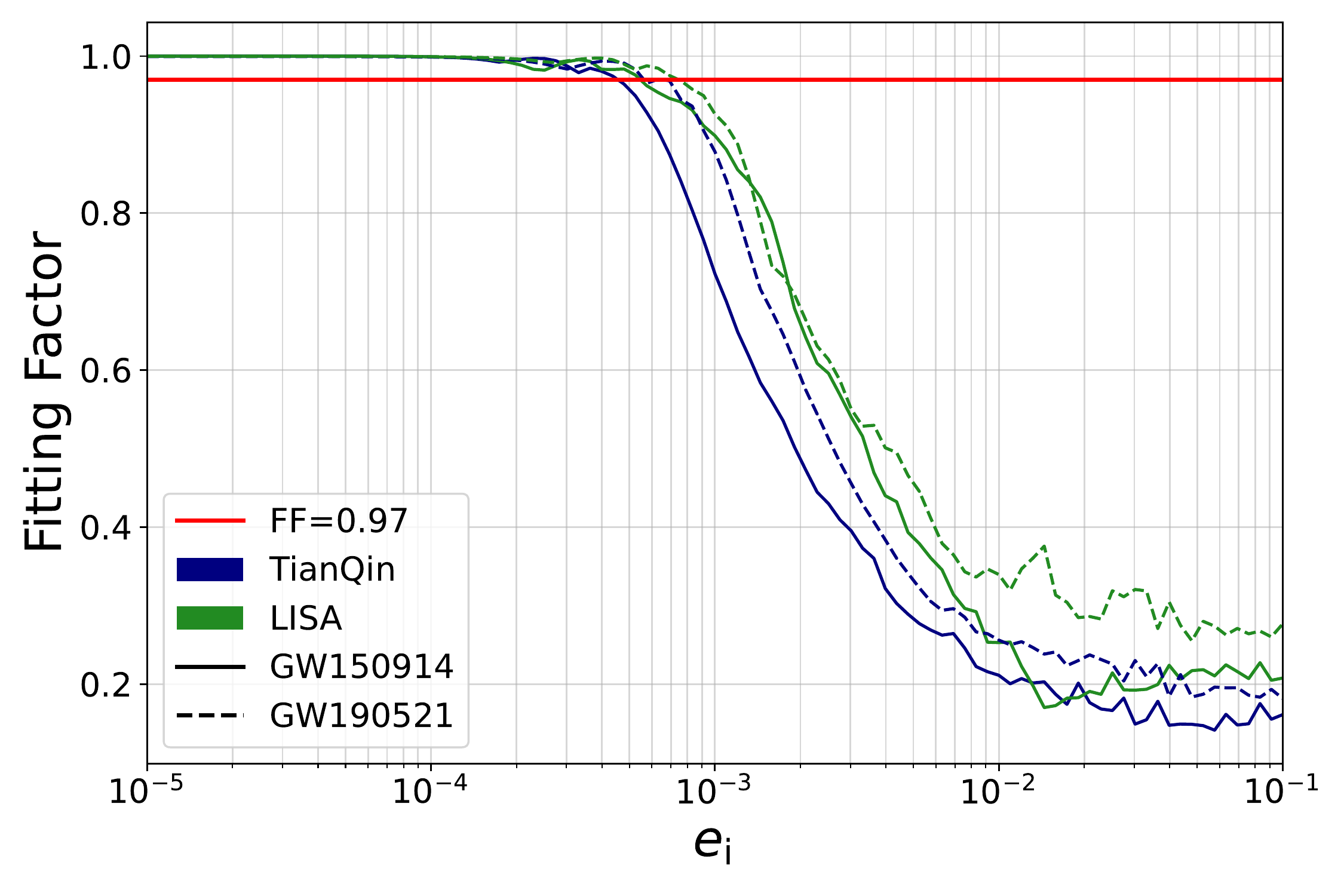}
	\caption{The fitting factor between a noneccentric template bank and a signal with different eccentricities. The blue(green) lines denote the banks of TianQin(LISA), the solid(dashed) lines correspond to the banks of a GW150914-like(GW190521-like) scenario. 
        }
	\label{ecc_bias}
\end{figure}

\section{Stochastic template bank generation}
If a signal has small eccentricity, it could be that a circular waveform would be sufficient to recover it. The question is, how small is small enough? We therefore use a noneccentric bank (i.e. banks of $\mathcal{M}$ in Table \ref{n_template}) and match it with an eccentric signal. In Fig \ref{ecc_bias}, we plot the fitting factor between the injected eccentric waveform and the template bank. As expected, the mismatch increases as eccentricity gets larger and we find that the eccentricity is distinguishable for TianQin/LISA when $ e_\mathrm{i} \gtrsim 5\times10^{-4} $. Many models, for example, dynamical interactions mechanisms~\cite{Kowalska2011,Nishizawa2017,Samsing2018}, predict larger initial eccentricity at $ \sim0.01\mathrm{Hz} $.
We also investigate the bias between the injected and recovered chirp mass when neglecting eccentricity, which increases from $ \lesssim10^{-6}M_\odot $ at $ e_\mathrm{i}=0 $ to $ \gtrsim10^{-3}M_\odot $ at $ e_\mathrm{i}=0.1 $. Such systematic bias could be even larger in the full parameter space. It is therefore necessary for searches to take eccentricity into account. 

\begin{table}[ht]
  \caption{Template bank sizes for GW150914- and GW190521-like events with different parameter spaces. }
	\centering
	\begin{tabular}{c|c|c|c}
		\hline
		\hline
		 & Parameter space & GW150914-like & GW190521-like \\
		\hline
		\hline
		\multirow{2}{*}{TianQin} & $ e_\mathrm{i}\in[0, 0.1] $                                     & 117202 & 49943 \\
		\cline{2-4}              & $ \mathcal{M}\in\mathcal{M}_0\pm10\sigma_\mathcal{M} $ & 3034   & 4250  \\
		\hline
		\hline
		\multirow{2}{*}{LISA} & $ e_\mathrm{i}\in[0, 0.1] $                                     & 100403 & 44867 \\
		\cline{2-4}           & $ \mathcal{M}\in\mathcal{M}_0\pm10\sigma_\mathcal{M} $ & 2070   & 3088  \\
		\hline
		\hline
	\end{tabular}
	\label{n_template}
\end{table}

\begin{figure}[ht]
	\centering
	\includegraphics[width=\linewidth]{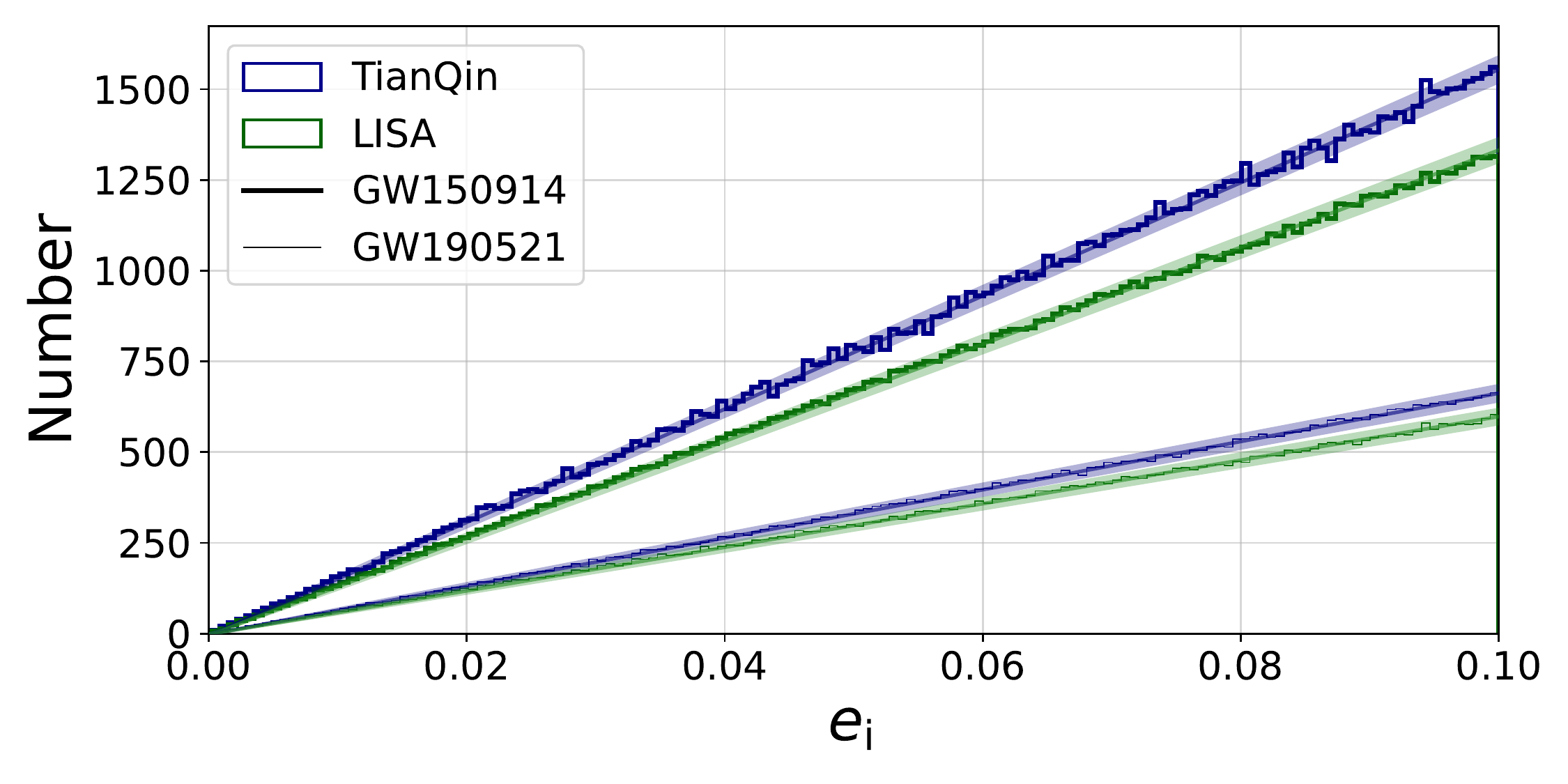}
        \caption{The distribution of the eccentricity in the archival search template bank. The shaded regions represent the $1\sigma$ Poisson fluctuation.}
	\label{distribution}
\end{figure}

In Table \ref{n_template} we show the size of the stochastic template banks, with different parameter spaces for both GW150914- and GW190521-like sources. 
We first assume that all the parameters (including chirp mass) are known exactly except for eccentricity, and thus generate a one-dimensional bank with $ e_\mathrm{i}\in[0, 0.1] $. The bank size is as large as $ \mathcal{O}(10^5) $ when only searching over eccentricity, and requires $ \lesssim80 $ core hours (and $ \lesssim 100$GB of memory). Therefore, for TianQin/LISA, we consider $ e_\mathrm{i} = 5\times10^{-4} $ ($ e_\mathrm{i}=0.1 $) as the smallest distinguishable eccentricity (the upper limit by the current computational cost), which corresponds to the red solid (dashed) line in Fig. \ref{citation_ecc}. 
Figure~\ref{distribution} shows the distribution of the eccentricity, which follows an $ e^2 $  cumulative distribution. 
It agrees with the theoretical estimate in previous studies~\cite{Nishizawa2016}, subject to Poisson fluctuation as indicated by the shaded region.
We then generate a one-dimensional bank covering only a range of chirp mass. Since the range is small, $\mathcal{M}$ appears to be uniformly distributed. 

As our current result fits well with a theoretical distribution for both eccentricity and chirp mass, we can give a reasonable estimation for the full 2D parameter space of the archival search. By assuming the 2D bank follows the same relationship as the 1D eccentric bank as the eccentricity range increases, the full 2D archival search banks are expected to have $ N_T \sim \mathcal{O}(10^8) $ templates, if we consider the maximal valid range for \texttt{EccentricFD}, i.e. $ e_\mathrm{i}\in[0,0.4] $, $ N_T $ will be up to $ \mathcal{O}(10^9) $.

To evaluate if we have overestimated the magnitude of 2D bank size due to any degeneracy between the eccentricity and the chirp mass~\cite{Favata2022,OShea2021,Lenon2020}, we generate a 2D bank for both detectors and for both sources. Restricted by the huge computational burden, we choose to verify the estimate through a bank within a smaller eccentricity range of $ e_\mathrm{i} \in {[0,0.001]} $.
All 2D banks have $N_T \sim \mathcal{O}(10^4)$, which is smaller but of the same order as the direct multiplication of bank sizes that are calculated separately in their parameter spaces. Such results do not change our magnitude estimation of the full 2D archival search bank size.
This indicates the challenge of computational cost: an example 2D bank with $ e_\mathrm{i} \in {[0,0.001]} $ includes 13372 templates, and would need $ \sim80\text{hr} $ for one core (and 18 GB of memory to cache waveforms) to generate.
By slicing the full parameter space along eccentricity and generating the 2D bank in parallel, 
a bank with $N_T \sim \mathcal{O}(10^8) $ needs $ \sim8\times10^5 $ core hours (and $ \sim 10^5$GB of memory).

\begin{figure}[ht]
	\centering
	\includegraphics[width=\linewidth]{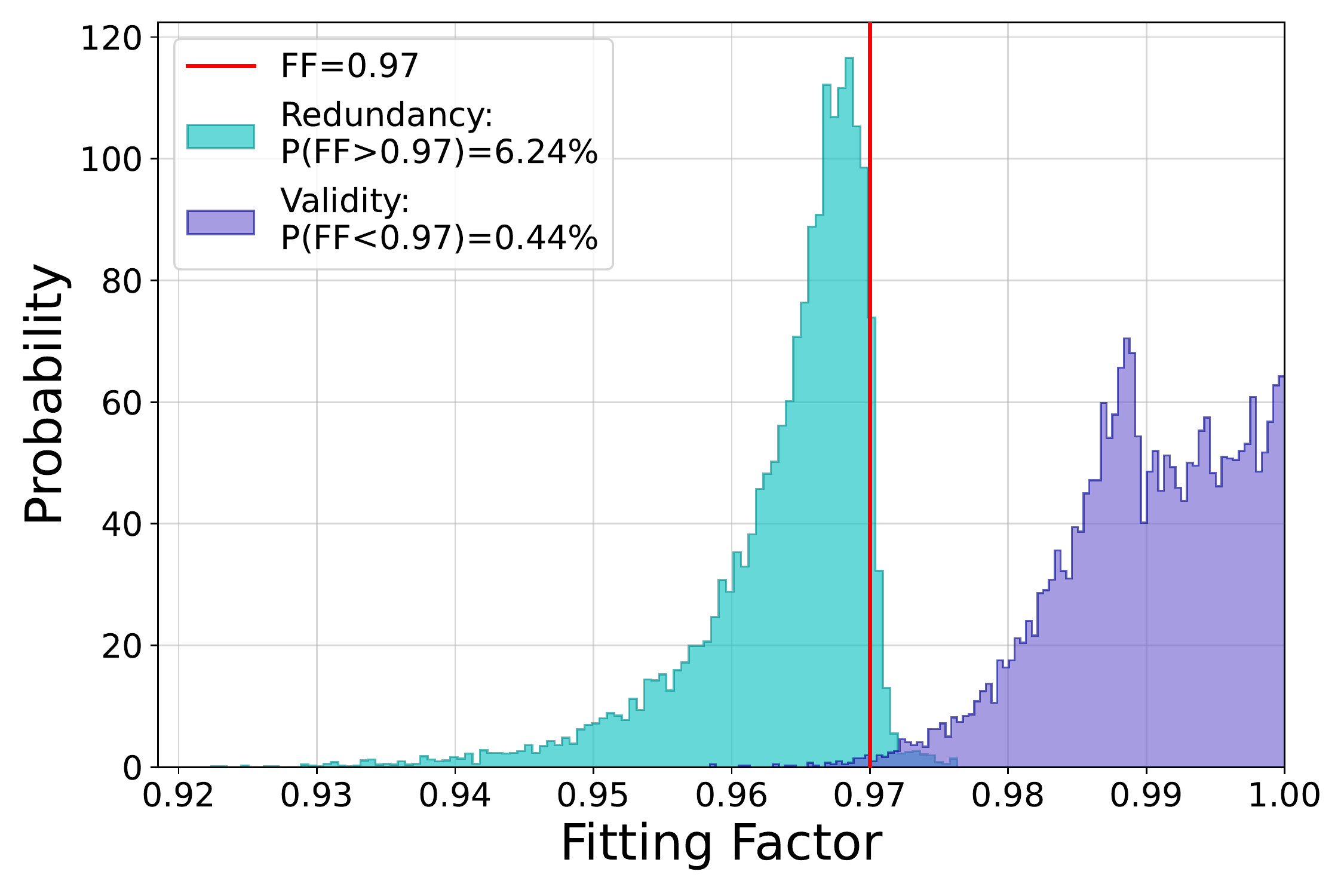}
        \caption{Validity and redundancy test of the example 2D template bank. The histogram in purple (cyan) shows the result of the validity test (redundancy test).  The vertical red line corresponds to the match criteria $ M=0.97 $.}
	\label{validity}
\end{figure}

To evaluate the performance of our template banks, we perform tests to quantify both the validity and redundancy. First we randomly generate 10,000 test waveforms with parameter values drawn from within the parameter space of the bank, and calculate the fitting factor for each waveform. If the bank is valid, all the test waveforms will have at least one template with which the match is larger than the minimal match threshold ($ M=0.97 $). In Fig. \ref{validity} we present the histogram of the fitting factor for the 10,000 injected waveforms. The red vertical line represents the threshold $M=0.97$, and we find that for almost all cases, the injected waveform has a $ {\rm FF} $ larger than 0.97, only 0.44\% of them fall lower than 0.97.

Then we move on to test the redundancy of the generated bank. We calculate the match between every template in the template bank. 
An ideal bank will have no redundancy, meaning the matches between all pairs of templates should be smaller than the minimal match threshold. 
In Fig \ref{validity}, following the validity test, for each template we present the histogram of the fitting factor, which is calculated on a bank that excludes the template itself. 
We find that only 6.22\% of all templates are redundant. 
This brings marginal extra computational cost. 

\section{Summary and discussion}
Numerous studies pointed out that the eccentricity of \acp{sBBH} will play a significant role in unveiling their origin.
In this paper, we demonstrate that the archival search of the \acp{sBBH} from space-based observatories is highly sensitive to the eccentricity.
Furthermore, for the first time, we successfully implement a \ac{GW} template bank generation process that includes eccentricity.

We generate one-dimensional template banks for either initial eccentricity or for chirp mass.
The upper limit of initial eccentricity at a system five years before merger is 0.1.
The range of chirp mass is determined by the estimation with the ground-based network.
By extrapolating the one-dimensional bank results, we conclude that a two-dimensional eccentric bank will comprise $ N_T \sim \mathcal{O}(10^{8-9}) $ templates, which is $ \sim\mathcal{O}(10^5) $ larger compared to the zero eccentricity case, and will require $ \sim\mathcal{O}(10^6) $ core hours [and $ \sim\mathcal{O}(10^5) $GB of memory] for the pipeline to generate it [also $ \sim\mathcal{O}(10^5) $ larger compared to noneccentric case].
This conclusion is verified by a small 2D bank, where the upper limit on the initial eccentricity is 0.001.
Constructing and filtering a template bank of $10^{8-9}$ waveforms will therefore be a challenging task, but it is not outside the scope of the expected computational facilities in the late
2030s, and could be further improved with additional optimization of the relevant software techniques.
Our work provides a practical solution to the realistic multiband GW observation scenario, with which one can determine the formation mechanism of \acp{sBBH} with successful archival searches.

It should be noted that we use a nonspinning eccentric waveform model in the paper. It is already known that spin effects are largely negligible during the inspiral\cite{Mangiagli2019} phase. However, in our technique, this is not a concern at all because the spin would already be constrained by ground-based facilities. Our space-based archival search would then just search a range of chirp mass and eccentricity values, using the measured black hole spins. It is important to note though that for both ground- and space-based detectors, more precise waveform models will be needed in the future to avoid potential systematic errors~\cite{Blanchet2014,Puerrer2020,RomeroShaw2021,RomeroShaw2023,Lyu2023}.

One caveat in the study is the duty cycle. We consider ET + dual CE for the ground-based detectors,
whereas in reality the duty cycle cannot reach 100\%; so the sky localization from realistic future networks might be worse than our calculation. 
Space-based observatories will also be limited by duty cycles\cite{Luo2016,Seoane2022}.
We leave the detailed calculation to future studies.

\section{Acknowledgments}
We are grateful to Xiang-Yu Lyu, En-Kun Li, Jian-dong Zhang and Shuai Liu for the helpful discussions.
This work has been supported by the Guangdong Major Project of Basic and Applied Basic Research (Grant No. 2019B030302001),  the Natural Science Foundation of China (Grant No. 12173104), and the National Key Research and Development Program of China (No. 2020YFC2201400). I.H. acknowledges support from the UK Space Agency through Grant No. ST/X002225/1.

\bibliographystyle{unsrtnat}
\bibliography{ref}

\begin{thebibliography}{68}
\providecommand{\natexlab}[1]{#1}
\providecommand{\url}[1]{\texttt{#1}}
\expandafter\ifx\csname urlstyle\endcsname\relax
  \providecommand{\doi}[1]{doi: #1}\else
  \providecommand{\doi}{doi: \begingroup \urlstyle{rm}\Url}\fi

\bibitem[{Orosz} et~al.(2011){Orosz}, {McClintock}, {Aufdenberg}, {Remillard},
  {Reid}, {Narayan}, and {Gou}]{Orosz2011}
Jerome~A. {Orosz}, Jeffrey~E. {McClintock}, Jason~P. {Aufdenberg}, Ronald~A.
  {Remillard}, Mark~J. {Reid}, Ramesh {Narayan}, and Lijun {Gou}.
\newblock {The Mass of the Black Hole in Cygnus X-1}.
\newblock \emph{Astrophys. J.}, 742\penalty0 (2):\penalty0 84, December 2011.
\newblock \doi{10.1088/0004-637X/742/2/84}.

\bibitem[Corral-Santana et~al.(2016)Corral-Santana, Casares, Munoz-Darias,
  Bauer, Martinez-Pais, and Russell]{CorralSantana2016}
Jesus~M. Corral-Santana, Jorge Casares, Teo Munoz-Darias, Franz~E. Bauer,
  Ignacio~G. Martinez-Pais, and David~M. Russell.
\newblock {BlackCAT: A catalogue of stellar-mass black holes in X-ray
  transients}.
\newblock \emph{Astron. Astrophys.}, 587:\penalty0 A61, 2016.
\newblock \doi{10.1051/0004-6361/201527130}.

\bibitem[{{\"O}zel} et~al.(2010){{\"O}zel}, {Psaltis}, {Narayan}, and
  {McClintock}]{Oezel2010}
Feryal {{\"O}zel}, Dimitrios {Psaltis}, Ramesh {Narayan}, and Jeffrey~E.
  {McClintock}.
\newblock {The Black Hole Mass Distribution in the Galaxy}.
\newblock \emph{Astrophys. J.}, 725\penalty0 (2):\penalty0 1918--1927, December
  2010.
\newblock \doi{10.1088/0004-637X/725/2/1918}.

\bibitem[Abbott et~al.(2016{\natexlab{a}})]{Abbott2016}
B.~P. Abbott et~al.
\newblock {Observation of Gravitational Waves from a Binary Black Hole Merger}.
\newblock \emph{Phys. Rev. Lett.}, 116\penalty0 (6):\penalty0 061102,
  2016{\natexlab{a}}.
\newblock \doi{10.1103/PhysRevLett.116.061102}.

\bibitem[Abbott et~al.(2016{\natexlab{b}})]{Abbott2016b}
B.~P. Abbott et~al.
\newblock {Astrophysical Implications of the Binary Black-Hole Merger
  GW150914}.
\newblock \emph{Astrophys. J. Lett.}, 818\penalty0 (2):\penalty0 L22,
  2016{\natexlab{b}}.
\newblock \doi{10.3847/2041-8205/818/2/L22}.

\bibitem[Abbott et~al.(2021)]{Abbott2021}
R.~Abbott et~al.
\newblock {GWTC-3: Compact Binary Coalescences Observed by LIGO and Virgo
  During the Second Part of the Third Observing Run}.
\newblock \emph{arXiv e-prints}, page arXiv:2111.03606, 11 2021.
\newblock URL \url{https://arxiv.org/abs/2111.03606}.

\bibitem[Nitz et~al.(2021)Nitz, Kumar, Wang, Kastha, Wu, Sch\"afer, Dhurkunde,
  and Capano]{Nitz2021}
Alexander~H. Nitz, Sumit Kumar, Yi-Fan Wang, Shilpa Kastha, Shichao Wu, Marlin
  Sch\"afer, Rahul Dhurkunde, and Collin~D. Capano.
\newblock {4-OGC: Catalog of gravitational waves from compact-binary mergers}.
\newblock \emph{arXiv e-prints}, art. arXiv:2112.06878, 12 2021.
\newblock \doi{10.48550/arXiv.2112.06878}.

\bibitem[Barack et~al.(2019)]{Barack2019}
Leor Barack et~al.
\newblock {Black holes, gravitational waves and fundamental physics: a
  roadmap}.
\newblock \emph{Class. Quant. Grav.}, 36\penalty0 (14):\penalty0 143001, 2019.
\newblock \doi{10.1088/1361-6382/ab0587}.

\bibitem[Abbott et~al.(2019)]{Abbott2019}
B.~P. Abbott et~al.
\newblock {Search for Eccentric Binary Black Hole Mergers with Advanced LIGO
  and Advanced Virgo during their First and Second Observing Runs}.
\newblock \emph{Astrophys. J.}, 883\penalty0 (2):\penalty0 149, 2019.
\newblock \doi{10.3847/1538-4357/ab3c2d}.

\bibitem[{Romero-Shaw} et~al.(2019){Romero-Shaw}, {Lasky}, and
  {Thrane}]{RomeroShaw2019}
Isobel~M. {Romero-Shaw}, Paul~D. {Lasky}, and Eric {Thrane}.
\newblock {Searching for Eccentricity: Signatures of Dynamical Formation in the
  First Gravitational-Wave Transient Catalogue of LIGO and Virgo}.
\newblock \emph{Mon. Not. Roy. Astron. Soc.}, 490\penalty0 (4):\penalty0
  5210--5216, December 2019.
\newblock \doi{10.1093/mnras/stz2996}.

\bibitem[{Wu} et~al.(2020){Wu}, {Cao}, and {Zhu}]{Wu2020}
Shichao {Wu}, Zhoujian {Cao}, and Zong-Hong {Zhu}.
\newblock {Measuring the eccentricity of binary black holes in GWTC-1 by using
  the inspiral-only waveform}.
\newblock \emph{Mon. Not. Roy. Astron. Soc.}, 495\penalty0 (1):\penalty0
  466--478, June 2020.
\newblock \doi{10.1093/mnras/staa1176}.

\bibitem[Abbott et~al.(2020)]{Abbott2020c}
R.~Abbott et~al.
\newblock {Properties and Astrophysical Implications of the 150 M$_\odot$
  Binary Black Hole Merger GW190521}.
\newblock \emph{Astrophys. J. Lett.}, 900\penalty0 (1):\penalty0 L13, 2020.
\newblock \doi{10.3847/2041-8213/aba493}.

\bibitem[Romero-Shaw et~al.(2020)Romero-Shaw, Lasky, Thrane, and
  Bustillo]{RomeroShaw2020}
Isobel~M. Romero-Shaw, Paul~D. Lasky, Eric Thrane, and Juan~Calderon Bustillo.
\newblock {GW190521: orbital eccentricity and signatures of dynamical formation
  in a binary black hole merger signal}.
\newblock \emph{Astrophys. J. Lett.}, 903\penalty0 (1):\penalty0 L5, 2020.
\newblock \doi{10.3847/2041-8213/abbe26}.

\bibitem[Gayathri et~al.(2022)Gayathri, Healy, Lange, O'Brien, Szczepanczyk,
  Bartos, Campanelli, Klimenko, Lousto, and O'Shaughnessy]{Gayathri2022}
V.~Gayathri, J.~Healy, J.~Lange, B.~O'Brien, M.~Szczepanczyk, Imre Bartos,
  M.~Campanelli, S.~Klimenko, C.~O. Lousto, and R.~O'Shaughnessy.
\newblock {Eccentricity estimate for black hole mergers with numerical
  relativity simulations}.
\newblock \emph{Nature Astron.}, 6\penalty0 (3):\penalty0 344--349, 2022.
\newblock \doi{10.1038/s41550-021-01568-w}.

\bibitem[Lower et~al.(2018)Lower, Thrane, Lasky, and Smith]{Lower2018}
Marcus~E. Lower, Eric Thrane, Paul~D. Lasky, and Rory Smith.
\newblock {Measuring eccentricity in binary black hole inspirals with
  gravitational waves}.
\newblock \emph{Phys. Rev. D}, 98\penalty0 (8):\penalty0 083028, 2018.
\newblock \doi{10.1103/PhysRevD.98.083028}.

\bibitem[Peters(1964)]{Peters1964}
P.~C. Peters.
\newblock {Gravitational Radiation and the Motion of Two Point Masses}.
\newblock \emph{Phys. Rev.}, 136:\penalty0 B1224--B1232, 1964.
\newblock \doi{10.1103/PhysRev.136.B1224}.

\bibitem[Rodriguez et~al.(2018)Rodriguez, Amaro-Seoane, Chatterjee, Kremer,
  Rasio, Samsing, Ye, and Zevin]{Rodriguez2018}
Carl~L. Rodriguez, Pau Amaro-Seoane, Sourav Chatterjee, Kyle Kremer,
  Frederic~A. Rasio, Johan Samsing, Claire~S. Ye, and Michael Zevin.
\newblock {Post-Newtonian Dynamics in Dense Star Clusters: Formation, Masses,
  and Merger Rates of Highly-Eccentric Black Hole Binaries}.
\newblock \emph{Phys. Rev. D}, 98\penalty0 (12):\penalty0 123005, 2018.
\newblock \doi{10.1103/PhysRevD.98.123005}.

\bibitem[Luo et~al.(2016)]{Luo2016}
Jun Luo et~al.
\newblock {TianQin: a space-borne gravitational wave detector}.
\newblock \emph{Class. Quant. Grav.}, 33\penalty0 (3):\penalty0 035010, 2016.
\newblock \doi{10.1088/0264-9381/33/3/035010}.

\bibitem[Amaro-Seoane et~al.(2017)Amaro-Seoane, Audley, Babak, Baker, Barausse,
  Bender, Berti, Binetruy, Born, Bortoluzzi, et~al.]{AmaroSeoane2017}
Pau Amaro-Seoane, Heather Audley, Stanislav Babak, John Baker, Enrico Barausse,
  Peter Bender, Emanuele Berti, Pierre Binetruy, Michael Born, Daniele
  Bortoluzzi, et~al.
\newblock {Laser Interferometer Space Antenna}.
\newblock \emph{arXiv e-prints}, art. arXiv:1702.00786, February 2017.

\bibitem[Nishizawa et~al.(2016)Nishizawa, Berti, Klein, and
  Sesana]{Nishizawa2016}
Atsushi Nishizawa, Emanuele Berti, Antoine Klein, and Alberto Sesana.
\newblock {eLISA eccentricity measurements as tracers of binary black hole
  formation}.
\newblock \emph{Phys. Rev. D}, 94\penalty0 (6):\penalty0 064020, 2016.
\newblock \doi{10.1103/PhysRevD.94.064020}.

\bibitem[Nishizawa et~al.(2017)Nishizawa, Sesana, Berti, and
  Klein]{Nishizawa2017}
Atsushi Nishizawa, Alberto Sesana, Emanuele Berti, and Antoine Klein.
\newblock {Constraining stellar binary black hole formation scenarios with
  eLISA eccentricity measurements}.
\newblock \emph{Mon. Not. Roy. Astron. Soc.}, 465\penalty0 (4):\penalty0
  4375--4380, 2017.
\newblock \doi{10.1093/mnras/stw2993}.

\bibitem[Chen and Amaro-Seoane(2017)]{Chen2017}
Xian Chen and Pau Amaro-Seoane.
\newblock {Revealing the formation of stellar-mass black hole binaries: The
  need for deci-Hertz gravitational wave observatories}.
\newblock \emph{Astrophys. J. Lett.}, 842\penalty0 (1):\penalty0 L2, 2017.
\newblock \doi{10.3847/2041-8213/aa74ce}.

\bibitem[{Liu} et~al.(2020){Liu}, {Hu}, {Zhang}, and {Mei}]{Liu2020}
Shuai {Liu}, Yi-Ming {Hu}, Jian-dong {Zhang}, and Jianwei {Mei}.
\newblock {Science with the TianQin observatory: Preliminary results on
  stellar-mass binary black holes}.
\newblock \emph{Phys. Rev. D}, 101\penalty0 (10):\penalty0 103027, May 2020.
\newblock \doi{10.1103/PhysRevD.101.103027}.

\bibitem[Buscicchio et~al.(2021)Buscicchio, Klein, Roebber, Moore, Gerosa,
  Finch, and Vecchio]{Buscicchio2021}
Riccardo Buscicchio, Antoine Klein, Elinore Roebber, Christopher~J. Moore,
  Davide Gerosa, Eliot Finch, and Alberto Vecchio.
\newblock {Bayesian parameter estimation of stellar-mass black-hole binaries
  with LISA}.
\newblock \emph{Phys. Rev. D}, 104\penalty0 (4):\penalty0 044065, 2021.
\newblock \doi{10.1103/PhysRevD.104.044065}.

\bibitem[{Klein} et~al.(2022){Klein}, {Pratten}, {Buscicchio}, {Schmidt},
  {Moore}, {Finch}, {Bonino}, {Thomas}, {Williams}, {Gerosa}, {McGee},
  {Nicholl}, and {Vecchio}]{Klein2022}
Antoine {Klein}, Geraint {Pratten}, Riccardo {Buscicchio}, Patricia {Schmidt},
  Christopher~J. {Moore}, Eliot {Finch}, Alice {Bonino}, Lucy~M. {Thomas},
  Natalie {Williams}, Davide {Gerosa}, Sean {McGee}, Matt {Nicholl}, and
  Alberto {Vecchio}.
\newblock {The last three years: multiband gravitational-wave observations of
  stellar-mass binary black holes}.
\newblock \emph{arXiv e-prints}, art. arXiv:2204.03423, April 2022.

\bibitem[Yunes et~al.(2009)Yunes, Arun, Berti, and Will]{Yunes2009}
Nicolas Yunes, K.~G. Arun, Emanuele Berti, and Clifford~M. Will.
\newblock Post-circular expansion of eccentric binary inspirals: Fourier-domain
  waveforms in the stationary phase approximation.
\newblock \emph{Phys. Rev. D}, 80:\penalty0 084001, Oct 2009.
\newblock \doi{10.1103/PhysRevD.80.084001}.
\newblock URL \url{https://link.aps.org/doi/10.1103/PhysRevD.80.084001}.

\bibitem[Kowalska et~al.(2011)Kowalska, Bulik, Belczynski, Dominik, and
  Gondek-Rosinska]{Kowalska2011}
I.~Kowalska, T.~Bulik, K.~Belczynski, M.~Dominik, and D.~Gondek-Rosinska.
\newblock {The eccentricity distribution of compact binaries}.
\newblock \emph{Astron. Astrophys.}, 527:\penalty0 A70, 2011.
\newblock \doi{10.1051/0004-6361/201015777}.

\bibitem[Breivik et~al.(2016)Breivik, Rodriguez, Larson, Kalogera, and
  Rasio]{Breivik2016}
Katelyn Breivik, Carl~L. Rodriguez, Shane~L. Larson, Vassiliki Kalogera, and
  Frederic~A. Rasio.
\newblock {Distinguishing Between Formation Channels for Binary Black Holes
  with LISA}.
\newblock \emph{Astrophys. J. Lett.}, 830\penalty0 (1):\penalty0 L18, 2016.
\newblock \doi{10.3847/2041-8205/830/1/L18}.

\bibitem[Rodriguez et~al.(2016)Rodriguez, Chatterjee, and Rasio]{Rodriguez2016}
Carl~L. Rodriguez, Sourav Chatterjee, and Frederic~A. Rasio.
\newblock {Binary Black Hole Mergers from Globular Clusters: Masses, Merger
  Rates, and the Impact of Stellar Evolution}.
\newblock \emph{Phys. Rev. D}, 93\penalty0 (8):\penalty0 084029, 2016.
\newblock \doi{10.1103/PhysRevD.93.084029}.

\bibitem[Samsing and D'Orazio(2018)]{Samsing2018}
Johan Samsing and Daniel~J. D'Orazio.
\newblock {Black Hole Mergers From Globular Clusters Observable by LISA I:
  Eccentric Sources Originating From Relativistic $N$-body Dynamics}.
\newblock \emph{Mon. Not. Roy. Astron. Soc.}, 481\penalty0 (4):\penalty0
  5445--5450, 2018.
\newblock \doi{10.1093/mnras/sty2334}.

\bibitem[Antonini and Perets(2012)]{Antonini2012}
Fabio Antonini and Hagai~B. Perets.
\newblock {Secular evolution of compact binaries near massive black holes:
  Gravitational wave sources and other exotica}.
\newblock \emph{Astrophys. J.}, 757:\penalty0 27, 2012.
\newblock \doi{10.1088/0004-637X/757/1/27}.

\bibitem[Antonini et~al.(2017)Antonini, Toonen, and Hamers]{Antonini2017}
Fabio Antonini, Silvia Toonen, and Adrian~S. Hamers.
\newblock {Binary black hole mergers from field triples: properties, rates and
  the impact of stellar evolution}.
\newblock \emph{Astrophys. J.}, 841\penalty0 (2):\penalty0 77, 2017.
\newblock \doi{10.3847/1538-4357/aa6f5e}.

\bibitem[Zevin et~al.(2019)Zevin, Samsing, Rodriguez, Haster, and
  Ramirez-Ruiz]{Zevin2019}
Michael Zevin, Johan Samsing, Carl Rodriguez, Carl-Johan Haster, and Enrico
  Ramirez-Ruiz.
\newblock {Eccentric Black Hole Mergers in Dense Star Clusters: The Role of
  Binary\textendash{}Binary Encounters}.
\newblock \emph{Astrophys. J.}, 871\penalty0 (1):\penalty0 91, 2019.
\newblock \doi{10.3847/1538-4357/aaf6ec}.

\bibitem[Samsing et~al.(2022)Samsing, Bartos, D'Orazio, Haiman, Kocsis, Leigh,
  Liu, Pessah, and Tagawa]{Samsing2022}
J.~Samsing, I.~Bartos, D.~J. D'Orazio, Z.~Haiman, B.~Kocsis, N.~W.~C. Leigh,
  B.~Liu, M.~E. Pessah, and H.~Tagawa.
\newblock {AGN as potential factories for eccentric black hole mergers}.
\newblock \emph{Nature}, 603\penalty0 (7900):\penalty0 237--240, 2022.
\newblock \doi{10.1038/s41586-021-04333-1}.

\bibitem[Xuan et~al.(2023)Xuan, Naoz, and Chen]{Xuan2023}
Zeyuan Xuan, Smadar Naoz, and Xian Chen.
\newblock {Detecting accelerating eccentric binaries in the LISA band}.
\newblock \emph{Phys. Rev. D}, 107\penalty0 (4):\penalty0 043009, 2023.
\newblock \doi{10.1103/PhysRevD.107.043009}.

\bibitem[Sun et~al.(2015)Sun, Cao, Wang, and Yeh]{Sun2015}
Baosan Sun, Zhoujian Cao, Yan Wang, and Hsien-Chi Yeh.
\newblock {Parameter estimation of eccentric inspiraling compact binaries using
  an enhanced post circular model for ground-based detectors}.
\newblock \emph{Phys. Rev. D}, 92\penalty0 (4):\penalty0 044034, 2015.
\newblock \doi{10.1103/PhysRevD.92.044034}.

\bibitem[Yang et~al.(2022)Yang, Cai, Cao, and Lee]{Yang2022}
Tao Yang, Rong-Gen Cai, Zhoujian Cao, and Hyung~Mok Lee.
\newblock {Eccentricity of Long Inspiraling Compact Binaries Sheds Light on
  Dark Sirens}.
\newblock \emph{Phys. Rev. Lett.}, 129\penalty0 (19):\penalty0 191102, 2022.
\newblock \doi{10.1103/PhysRevLett.129.191102}.

\bibitem[Saini et~al.(2022)Saini, Favata, and Arun]{Saini2022}
Pankaj Saini, Marc Favata, and K.~G. Arun.
\newblock {Systematic bias on parametrized tests of general relativity due to
  neglect of orbital eccentricity}.
\newblock \emph{Phys. Rev. D}, 106\penalty0 (8):\penalty0 084031, 2022.
\newblock \doi{10.1103/PhysRevD.106.084031}.

\bibitem[Bhat et~al.(2023)Bhat, Saini, Favata, and Arun]{Bhat2023}
Sajad~A. Bhat, Pankaj Saini, Marc Favata, and K.~G. Arun.
\newblock {Systematic bias on the inspiral-merger-ringdown consistency test due
  to neglect of orbital eccentricity}.
\newblock \emph{Phys. Rev. D}, 107\penalty0 (2):\penalty0 024009, 2023.
\newblock \doi{10.1103/PhysRevD.107.024009}.

\bibitem[Babak et~al.(2013)]{Babak2013}
S.~Babak et~al.
\newblock {Searching for gravitational waves from binary coalescence}.
\newblock \emph{Phys. Rev. D}, 87\penalty0 (2):\penalty0 024033, 2013.
\newblock \doi{10.1103/PhysRevD.87.024033}.

\bibitem[Dal~Canton and Harry(2017)]{DalCanton2017}
Tito Dal~Canton and Ian~W. Harry.
\newblock {Designing a template bank to observe compact binary coalescences in
  Advanced LIGO's second observing run}.
\newblock \emph{arXiv e-prints}, 5 2017.

\bibitem[{Moore} et~al.(2019){Moore}, {Gerosa}, and {Klein}]{Moore2019a}
Christopher~J. {Moore}, Davide {Gerosa}, and Antoine {Klein}.
\newblock {Are stellar-mass black-hole binaries too quiet for LISA?}
\newblock \emph{Mon. Not. Roy. Astron. Soc.}, 488\penalty0 (1):\penalty0
  L94--L98, September 2019.
\newblock \doi{10.1093/mnrasl/slz104}.

\bibitem[Sesana(2016)]{Sesana2016}
Alberto Sesana.
\newblock {Prospects for Multiband Gravitational-Wave Astronomy after
  GW150914}.
\newblock \emph{Phys. Rev. Lett.}, 116\penalty0 (23):\penalty0 231102, 2016.
\newblock \doi{10.1103/PhysRevLett.116.231102}.

\bibitem[{Wong} et~al.(2018){Wong}, {Kovetz}, {Cutler}, and {Berti}]{Wong2018}
Kaze W.~K. {Wong}, Ely~D. {Kovetz}, Curt {Cutler}, and Emanuele {Berti}.
\newblock {Expanding the LISA Horizon from the Ground}.
\newblock \emph{Phys. Rev. Lett.}, 121\penalty0 (25):\penalty0 251102, December
  2018.
\newblock \doi{10.1103/PhysRevLett.121.251102}.

\bibitem[{Ewing} et~al.(2021){Ewing}, {Sachdev}, {Borhanian}, and
  {Sathyaprakash}]{Ewing2021}
Becca {Ewing}, Surabhi {Sachdev}, Ssohrab {Borhanian}, and B.~S.
  {Sathyaprakash}.
\newblock {Archival searches for stellar-mass binary black holes in LISA data}.
\newblock \emph{Phys. Rev. D}, 103\penalty0 (2):\penalty0 023025, January 2021.
\newblock \doi{10.1103/PhysRevD.103.023025}.

\bibitem[Punturo et~al.(2010)Punturo, Abernathy, Acernese, Allen, Andersson,
  Arun, Barone, Barr, Barsuglia, Beker, et~al.]{Punturo2010}
M~Punturo, M~Abernathy, F~Acernese, B~Allen, Nils Andersson, K~Arun, F~Barone,
  B~Barr, M~Barsuglia, M~Beker, et~al.
\newblock {The Einstein Telescope: a third-generation gravitational wave
  observatory}.
\newblock \emph{Class. Quant. Grav.}, 27\penalty0 (19):\penalty0 194002,
  October 2010.
\newblock \doi{10.1088/0264-9381/27/19/194002}.

\bibitem[Reitze et~al.(2019)Reitze, Adhikari, Ballmer, Barish, Barsotti,
  Billingsley, Brown, Chen, Coyne, Eisenstein, et~al.]{Reitze2019}
David Reitze, Rana~X Adhikari, Stefan Ballmer, Barry Barish, Lisa Barsotti,
  GariLynn Billingsley, Duncan~A Brown, Yanbei Chen, Dennis Coyne, Robert
  Eisenstein, et~al.
\newblock {Cosmic Explorer: The U.S. Contribution to Gravitational-Wave
  Astronomy beyond LIGO}.
\newblock In \emph{Bulletin of the American Astronomical Society}, volume~51,
  page~35, September 2019.

\bibitem[Cho(2022)]{Cho2022}
Hee-Suk Cho.
\newblock {Improvement of the parameter measurement accuracy by the
  third-generation gravitational wave detector Einstein Telescope}.
\newblock \emph{Class. Quant. Grav.}, 39\penalty0 (8):\penalty0 085006, 2022.
\newblock \doi{10.1088/1361-6382/ac5b31}.

\bibitem[Huerta et~al.(2014)Huerta, Kumar, McWilliams, O'Shaughnessy, and
  Yunes]{Huerta2014}
E.~A. Huerta, Prayush Kumar, Sean~T. McWilliams, Richard O'Shaughnessy, and
  Nicol\'as Yunes.
\newblock Accurate and efficient waveforms for compact binaries on eccentric
  orbits.
\newblock \emph{Phys. Rev. D}, 90:\penalty0 084016, Oct 2014.
\newblock \doi{10.1103/PhysRevD.90.084016}.
\newblock URL \url{https://link.aps.org/doi/10.1103/PhysRevD.90.084016}.

\bibitem[{LIGO Scientific Collaboration}(2018)]{LSC2018}
{LIGO Scientific Collaboration}.
\newblock {LIGO} {A}lgorithm {L}ibrary - {LALS}uite.
\newblock free software (GPL), 2018.

\bibitem[London et~al.(2018)London, Khan, Fauchon-Jones, Garc\'\i{}a, Hannam,
  Husa, Jim\'enez-Forteza, Kalaghatgi, Ohme, and Pannarale]{London2018}
Lionel London, Sebastian Khan, Edward Fauchon-Jones, Cecilio Garc\'\i{}a, Mark
  Hannam, Sascha Husa, Xisco Jim\'enez-Forteza, Chinmay Kalaghatgi, Frank Ohme,
  and Francesco Pannarale.
\newblock {First higher-multipole model of gravitational waves from spinning
  and coalescing black-hole binaries}.
\newblock \emph{Phys. Rev. Lett.}, 120\penalty0 (16):\penalty0 161102, 2018.
\newblock \doi{10.1103/PhysRevLett.120.161102}.

\bibitem[Lenon et~al.(2021)Lenon, Brown, and Nitz]{Lenon2021}
Amber~K. Lenon, Duncan~A. Brown, and Alexander~H. Nitz.
\newblock {Eccentric binary neutron star search prospects for Cosmic Explorer}.
\newblock \emph{Phys. Rev. D}, 104\penalty0 (6):\penalty0 063011, 2021.
\newblock \doi{10.1103/PhysRevD.104.063011}.

\bibitem[sba()]{sbank}
\url{https://github.com/gwastro/sbank}.

\bibitem[Babak(2008)]{Babak2008}
Stanislav Babak.
\newblock {Building a stochastic template bank for detecting massive black hole
  binaries}.
\newblock \emph{Class. Quant. Grav.}, 25:\penalty0 195011, 2008.
\newblock \doi{10.1088/0264-9381/25/19/195011}.

\bibitem[Harry et~al.(2009)Harry, Allen, and Sathyaprakash]{Harry2009}
Ian~W. Harry, Bruce Allen, and B.~S. Sathyaprakash.
\newblock {A Stochastic template placement algorithm for gravitational wave
  data analysis}.
\newblock \emph{Phys. Rev. D}, 80:\penalty0 104014, 2009.
\newblock \doi{10.1103/PhysRevD.80.104014}.
\newblock URL \url{https://link.aps.org/doi/10.1103/PhysRevD.80.104014}.

\bibitem[Creighton and Anderson(2011)]{Creighton2011}
Jolien D.~E. Creighton and Warren~G. Anderson.
\newblock \emph{Gravitational-wave physics and astronomy: An introduction to
  theory, experiment and data analysis}.
\newblock John Wiley \& Sons, 2011.

\bibitem[Marsat et~al.(2021)Marsat, Baker, and Canton]{Marsat2021}
Sylvain Marsat, John~G. Baker, and Tito~Dal Canton.
\newblock Exploring the bayesian parameter estimation of binary black holes
  with lisa.
\newblock \emph{Phys. Rev. D}, 103:\penalty0 083011, Apr 2021.
\newblock \doi{10.1103/PhysRevD.103.083011}.
\newblock URL \url{https://link.aps.org/doi/10.1103/PhysRevD.103.083011}.

\bibitem[Lyu et~al.(2023)Lyu, Li, and Hu]{Lyu2023}
Xiangyu Lyu, En-Kun Li, and Yi-Ming Hu.
\newblock {Parameter Estimation of Stellar Mass Binary Black Holes under the
  Network of TianQin and LISA}.
\newblock \emph{Phys. Rev. D}, 108\penalty0 (8):\penalty0 083023, 7 2023.
\newblock \doi{10.1103/PhysRevD.108.083023}.

\bibitem[Babak et~al.(2021)Babak, Petiteau, and Hewitson]{Babak2021}
Stanislav Babak, Antoine Petiteau, and Martin Hewitson.
\newblock {LISA Sensitivity and SNR Calculations}.
\newblock \emph{arXiv e-prints}, 8 2021.

\bibitem[{Favata} et~al.(2022){Favata}, {Kim}, {Arun}, {Kim}, and
  {Lee}]{Favata2022}
Marc {Favata}, Chunglee {Kim}, K.~G. {Arun}, JeongCho {Kim}, and Hyung~Won
  {Lee}.
\newblock {Constraining the orbital eccentricity of inspiralling compact binary
  systems with Advanced LIGO}.
\newblock \emph{Phys. Rev. D}, 105\penalty0 (2):\penalty0 023003, January 2022.
\newblock \doi{10.1103/PhysRevD.105.023003}.

\bibitem[O'Shea and Kumar(2021)]{OShea2021}
Eamonn O'Shea and Prayush Kumar.
\newblock {Correlations in parameter estimation of low-mass eccentric binaries:
  GW151226 \& GW170608}.
\newblock \emph{arXiv e-prints}, page arXiv:2107.07981, 7 2021.

\bibitem[Lenon et~al.(2020)Lenon, Nitz, and Brown]{Lenon2020}
Amber~K. Lenon, Alexander~H. Nitz, and Duncan~A. Brown.
\newblock {Measuring the eccentricity of GW170817 and GW190425}.
\newblock \emph{Mon. Not. Roy. Astron. Soc.}, 497\penalty0 (2):\penalty0
  1966--1971, 2020.
\newblock \doi{10.1093/mnras/staa2120}.

\bibitem[Mangiagli et~al.(2019)Mangiagli, Klein, Sesana, Barausse, and
  Colpi]{Mangiagli2019}
Alberto Mangiagli, Antoine Klein, Alberto Sesana, Enrico Barausse, and Monica
  Colpi.
\newblock {Post-Newtonian phase accuracy requirements for stellar black hole
  binaries with LISA}.
\newblock \emph{Phys. Rev. D}, 99\penalty0 (6):\penalty0 064056, 2019.
\newblock \doi{10.1103/PhysRevD.99.064056}.

\bibitem[Blanchet(2014)]{Blanchet2014}
Luc Blanchet.
\newblock {Gravitational Radiation from Post-Newtonian Sources and Inspiralling
  Compact Binaries}.
\newblock \emph{Living Rev. Rel.}, 17:\penalty0 2, 2014.
\newblock \doi{10.12942/lrr-2014-2}.

\bibitem[P\"urrer and Haster(2020)]{Puerrer2020}
Michael P\"urrer and Carl-Johan Haster.
\newblock {Gravitational waveform accuracy requirements for future ground-based
  detectors}.
\newblock \emph{Phys. Rev. Res.}, 2\penalty0 (2):\penalty0 023151, 2020.
\newblock \doi{10.1103/PhysRevResearch.2.023151}.

\bibitem[Romero-Shaw et~al.(2021)Romero-Shaw, Lasky, and
  Thrane]{RomeroShaw2021}
Isobel~M. Romero-Shaw, Paul~D. Lasky, and Eric Thrane.
\newblock {Signs of Eccentricity in Two Gravitational-wave Signals May Indicate
  a Subpopulation of Dynamically Assembled Binary Black Holes}.
\newblock \emph{Astrophys. J. Lett.}, 921\penalty0 (2):\penalty0 L31, 2021.
\newblock \doi{10.3847/2041-8213/ac3138}.

\bibitem[Romero-Shaw et~al.(2023)Romero-Shaw, Gerosa, and
  Loutrel]{RomeroShaw2023}
Isobel~M. Romero-Shaw, Davide Gerosa, and Nicholas Loutrel.
\newblock {Eccentricity or spin precession? Distinguishing subdominant effects
  in gravitational-wave data}.
\newblock \emph{Mon. Not. Roy. Astron. Soc.}, 519\penalty0 (4):\penalty0
  5352--5357, 2023.
\newblock \doi{10.1093/mnras/stad031}.

\bibitem[Seoane et~al.(2022)]{Seoane2022}
Pau~Amaro Seoane et~al.
\newblock {The effect of mission duration on LISA science objectives}.
\newblock \emph{Gen. Rel. Grav.}, 54\penalty0 (1):\penalty0 3, 2022.
\newblock \doi{10.1007/s10714-021-02889-x}.

\end{thebibliography}

\end{document}